\documentclass[times, 10pt,twocolumn]{article}
\usepackage{times}
\usepackage{latex8}
\usepackage{cite}
\usepackage{graphicx}
\usepackage{url}
\newcommand{\ie}{i.\,e.\ }
\newcommand{\eg}{e.\,g.\ }
\usepackage{url} 
\usepackage{graphicx}
\usepackage{latexsym}
\usepackage{cite}

\newcommand{\qmm}{\textsc{Qmm}\xspace}

\newcommand{\impact}[4]{\textsf{\small [#1$\,\vert\,$\uppercase{{\scriptsize#2}}] 
$\stackrel{#4}{\longrightarrow}$ [#3]}\xspace}

\newcommand{\impactp}[3]{\impact{#1}{#2}{#3}{+}}
\newcommand{\impactn}[3]{\impact{#1}{#2}{#3}{-}}

\newcommand{\fact}[2]{\textsf{\small [#1$\,\vert\,$\uppercase{{\scriptsize#2}}]}}
\newcommand{\entity}[1]{\textsf{\small #1}}
\newcommand{\attribute}[1]{\textsf{\small\uppercase{{\scriptsize#1}}}}

\usepackage{xspace}

\begin{document}

\title{An Activity-Based Quality Model for Maintainability}

\author{F.~Deissenboeck, S.~Wagner, M.~Pizka\\
Institut f\"ur Informatik\\
Technische Universit\"at M\"unchen\\
Boltzmannstr. 3\\
85748 Garching b.~M\"unchen, Germany\\
\{deissenb,wagnerst,pizka\}@in.tum.de
\and 
S.~Teuchert, J.-F.~Girard\\
MAN Nutzfahrzeuge AG\\
Elektronik Regelungs- und Steuerungssysteme (TSE)\\
Dachauer Strasse 667\\ 80995 M\"unchen, Germany\\
\{stefan.teuchert,jean-francois.girard\}@man.eu
}

\maketitle
\thispagestyle{empty}

\begin{abstract}

Maintainability is a key quality attribute of successful software systems. 
However, its management in practice is still problematic. Currently,
there is no comprehensive basis for assessing and improving the
maintainability of software systems. Quality models have been proposed to
solve this problem.
Nevertheless, existing approaches do not explicitly 
take into account the maintenance activities, that largely determine the
software maintenance effort. This paper proposes a 2-dimensional model of 
maintainability that explicitly associates system properties with the 
activities carried out during maintenance. The separation of activities and 
properties facilitates the identification of sound quality criteria and 
allows to reason about their interdependencies. 
This transforms the quality model into a structured and comprehensive quality
knowledge base that is usable in industrial project environments. For example,
review guidelines can be generated from it. The model is based on an 
explicit quality metamodel that supports its systematic construction 
and fosters preciseness as well as completeness. An 
industrial case study demonstrates the applicability of the model for the 
evaluation of the maintainability of Matlab Simulink models that are 
frequently used in model-based development of embedded systems.

\end{abstract}

\Section{Introduction}

Virtually any software dependent organization has a vital interest in 
reducing its spending for software maintenance activities.   In addition to 
financial savings, for many organizations, the time needed to complete a 
software maintenance task largely determines their ability to adapt their 
business processes to changing market situations or to implement innovative 
products and services. That is to say that with the present yet increasing 
dependency on large scale software systems, the ability to change existing 
software in a timely and economical manner becomes critical for 
numerous enterprises of diverse branches. The term most frequently 
associated with this is \emph{maintainability}. But what is 
maintainability? 
An often cited definition is:
\emph{The effort needed to
make specified modifications to a component implementation}.\footnote{SEI
Open Systems Glossary
(\url{http://www.sei.cmu.edu/opensystems/glossary.html)}}

This nicely illustrates that the desire for \emph{high maintainability} is 
really a desire for \emph{low maintenance efforts}. However, current 
approaches to assess and improve maintainability fail to explicitly take 
into account the cost factor that largely determines software maintenance 
efforts: the \emph{activities} performed on the system or more precisely, 
the associated personnel costs. Considering the diverse nature of 
activities, such as ``problem understanding'' and ``testing'', it becomes 
evident, that the criteria that actually influence the maintenance effort 
are numerous and diverse. Psychological effects, such as the \emph{broken 
window}\,\cite{1982_wilson_broken_window} deserve just as much attention as 
organizational issues (e.\,g.~personnel turnover) or properties of the 
code.  Any of 
these aspects may have a significant and vastly independent impact on the 
future maintenance effort.

We regard the omission of activities as a serious flaw not only due to the 
activities' major importance for the overall maintenance effort but also 
because the activities provide a natural criterion for the decomposition of 
maintainability that many existing approaches lack.

\paragraph{Problem} Although maintainability is a key quality attribute of 
large software systems, existing approaches to model maintainability have 
not created a common understanding of the factors influencing 
maintainability and their interrelations. Hence, no comprehensive basis 
for assessing and improving the maintainability of large software systems
has been established so far.

Typically, existing models exhibit at least one of the following problems: 
First, they do not decompose the attributes and criteria to a level that is 
suitable for an actual assessment. Second, these models tend to omit 
the rationale behind the required properties of the system. Third, 
existing models often use heterogeneous decomposition dimensions, \eg the 
required criteria mix properties of the system with properties of the 
activities carried out on the system.

The first problem constrains the use of these models as the basis for 
analyses. The second one makes it difficult to describe impacts precisely 
and therefore to convince developers to use it. The third problem leads to 
inconsistent models and hampers the revelation of omissions and 
inconsistencies in these models. 

\paragraph{Contribution} This paper proposes a 2-dimensional model of 
maintainability that explicitly associates system properties with the 
activities carried out during maintenance and thereby facilitates a 
structured decomposition of maintainability. The separation of activities 
and properties facilitates the identification of sound quality criteria and 
allows to reason about their interdependencies. As the activities are the 
main cost factor in software maintenance, we consider this separation a 
first step towards the ultimate goal of a truly economically justified 
practice of maintainability engineering. The model is based on an explicit 
quality metamodel that supports a systematic construction of a 
maintainability model and fosters preciseness as well as completeness.

Next to the ability to explicitly describe the impact of system properties 
on the maintenance activities, several additional benefits can be realized 
by using the model:

\begin{enumerate}
  
  \item The model provides a central storage of quality definition, 
  comparable to a knowledge base, that serves as a basis for the automatic 
  generation of guideline documents for specific maintenance tasks as well 
  as for the analysis of system artifacts.
  
  \item The model allows to reveal omissions and contradictions in current 
  models and guidelines.
  
\end{enumerate}
We demonstrate the applicability of the 2-dimensional model in a case study 
undertaken with MAN Nutzfahrzeuge, a German supplier of commercial 
vehicles and transport systems. Here we created a 
comprehensive model of the maintainability of Matlab Simulink models 
that are frequently used in model-based development of 
embedded systems. The study lead to the inclusion of the model into the MAN 
standard development process.

\Section{Related Work}
To be read conveniently the related work is categorized as guidelines-based 
approaches, metrics-based approaches, quality models and process-based 
approaches.

\paragraph{Guidelines} A commonly applied practice are guidelines that 
state what developers should do and what they should not do in order to 
improve the quality of software artifacts. Such guidelines are usually 
composed by the software-developing companies itself or by tool providers, 
\eg the Java Coding Conventions provided by Sun 
Microsystems~\cite{1999_sun_java_conventions}. 

Unfortunately, such guidelines typically do not achieve the desired effect 
as developers often read them once, tuck them away at the bottom of a 
drawer and follow them in a sporadic manner only. According to our 
experience~\cite{2006_broym_maintainability}, this is often due to the fact 
that guidelines fail to motivate the required practices or provide very 
generic explanations, \eg ``Respecting the guideline ensures readable 
models'' in \cite{2001_maab_guideline}.
Justification could be provided by explaining how 
conformance/non-conformance to guidelines effects maintenance activities
and thereby maintenance effort.

In addition to this, guidelines are often not followed simply because it is 
not checked if they are followed or not. This is all the more unfortunate 
as for some guidelines rules compliance could be assessed automatically.

\paragraph{Metrics-based Approaches}

Several groups proposed me\-trics-based methods to measure attributes of
software systems which are believed to affect
maintenance, \eg~\cite{1984_bernsg_maintainability_assessment,
  1994_colemand_maintainability_metrics}. Typically, these methods use a
set of well-known metrics like \emph{lines of code}, Halstead volume
\cite{1977_halsteadm_software_science}, or McCabe's Cyclomatic
Complexity\,\cite{1976_mccabej_complexity_measure} and combine them into a
single value, called \emph{maintainability index} by means of statistically
determined weights.

Although such indices may indeed often expose a correlation with subjective
impressions and economic facts of a software system, they still suffer
from serious shortcomings. First, they do not explain in which way system
properties influence the maintenance activities and thereby the overall
maintenance efforts. This makes it hard to convey their findings to the
developers.

Second, they focus on properties which can be measured automatically by
analyzing source code and thereby limit themselves to syntactic aspects.
However, many essential quality issues, such as the usage of
appropriate data structures and meaningful documentation, are semantic in
nature and can inherently not be analyzed automatically.

Because of this, most known metrics, such as the Cyclomatic Complexity, are 
neither sufficient nor necessary to indicate a quality defect. Therefore, 
individual metrics or simple indices provide only a poor basis for 
effective quality assessments.

\paragraph{Quality Modeling}

A promising approach developed for software quality in general are 
\emph{quality models} which aim at describing complex quality criteria by 
breaking them down into more manageable sub-criteria. Such models are 
designed in a tree-like fashion with abstract quality attributes like 
\emph{maintainability} or \emph{reliability} at the top and more concrete 
ones like \emph{analyzability} or \emph{changeability} on lower levels. The 
leaf factors are ideally detailed enough to be assessed with software 
metrics.  This method is frequently called the decompositional or 
\emph{Factor-Criteria-Metric} (FCM) approach and was first used by 
McCall\,\cite{1977_mccallj_quality_factors} and 
Boehm\,\cite{1978_boehmb_software_quality}.

Nevertheless, these and more recent approaches 
like~\cite{1992_omanp_maintainability,  
1995_dromeyr_product_quality, 2004_marinescur_oo_analysis, 
2003_iso_standard_9126_1} have failed to establish a broadly acceptable 
basis for quality assessments so far. We believe this is due to the lack 
of a clearly defined decomposition criterion that leads to a ``somewhat 
arbitrary selection of characteristics and 
sub-characteristics''~\cite{1997_kitchenhamb_squid_quality_model, 
1996_kitchenhamb_quality_elusive_target}. Moreover, we see  their fixed 
number of model levels as a problem. For example, FCM's 3 level structure 
is inadequate. Breaking down high level goals like \emph{maintainability}
into assessable properties leads to a loose connection between criteria
and metrics.

Similar to other approaches, quality models do usually not model 
the maintenance activities explicitly. Hence, they are not directly capable of 
explaining how system properties influence the maintenance effort.

\paragraph{Processes and Process Models} Organizational issues are 
typically covered by process-based approaches to software quality like the 
ISO~9000 standards or CMM\,\cite{1995_paulkm_cmm}. Unfortunately, there is 
the widely disputed misconception, that good processes automatically 
guarantee high quality 
products\,\cite{1996_kitchenhamb_quality_elusive_target}. Of course, 
processes are of high importance and they determine reproducibility of the 
development process.  However, the quality of the outcome still strongly 
depends on the actual criteria, skills, and tools used during development. 

\paragraph{Discussion}

Garvin describes in \cite{1984_garvind_product_quality} quality as a ``complex 
and multifaceted concept'' and discusses different but equally valid 
perspectives on the subject. We consider these different perspectives and the 
inherent complexity of quality itself the reason for the variety of different 
approaches discussed above. Consequently, there is an abundance of further 
highly valuable work on software quality in general and maintainability in 
particular that we do not explicitly mention here, as it is either out-of-scope 
or does not fundamentally differ from the work already mentioned. 

Although this has been and continues to be a very active field of research, 
we argue that 
existing approaches to assess and improve software maintainability generally 
suffer from one or more of the following shortcomings:

\begin{enumerate}

\item \emph{Assessability.} Most quality models contain a number of 
criteria that are too coarse-grained to be assessed directly. 

\item \emph{Justification.} Additionally, most existing quality models fail 
to give a detailed account of the impact that specific criteria (or 
metrics) have on software maintenance. 

\item \emph{Homogeneity.} Due to the lack of a consistent criterion of 
decomposition  most existing models exhibit inhomogeneous sets of quality 
criteria. 

\item \emph{Operationalization.} Most times, quality models are expressed
in prose and graphics only. They accompany the development process
in the form of documents but are not made an integral artifact that
is tightly coupled with the quality assurance activities.

\end{enumerate}

\Section{Maintainability Model\label{sec:qmm}}

To address the problems of quality models described in the previous 
section we developed a novel 2-dimensional quality model. The initial 
version of the model was developed in the context of a commercial project 
in the field of telecommunication~\cite{2006_broym_maintainability}. As the 
analyzed system was large (3.5 MLOC\footnote{million lines of code} C++, 
COBOL, Java), 15 years old and under active maintenance with 150 change 
requests per year it was well suited for an application of our quality 
model. 

\begin{figure}[h]
\begin{center}
\includegraphics[width=0.99\linewidth]{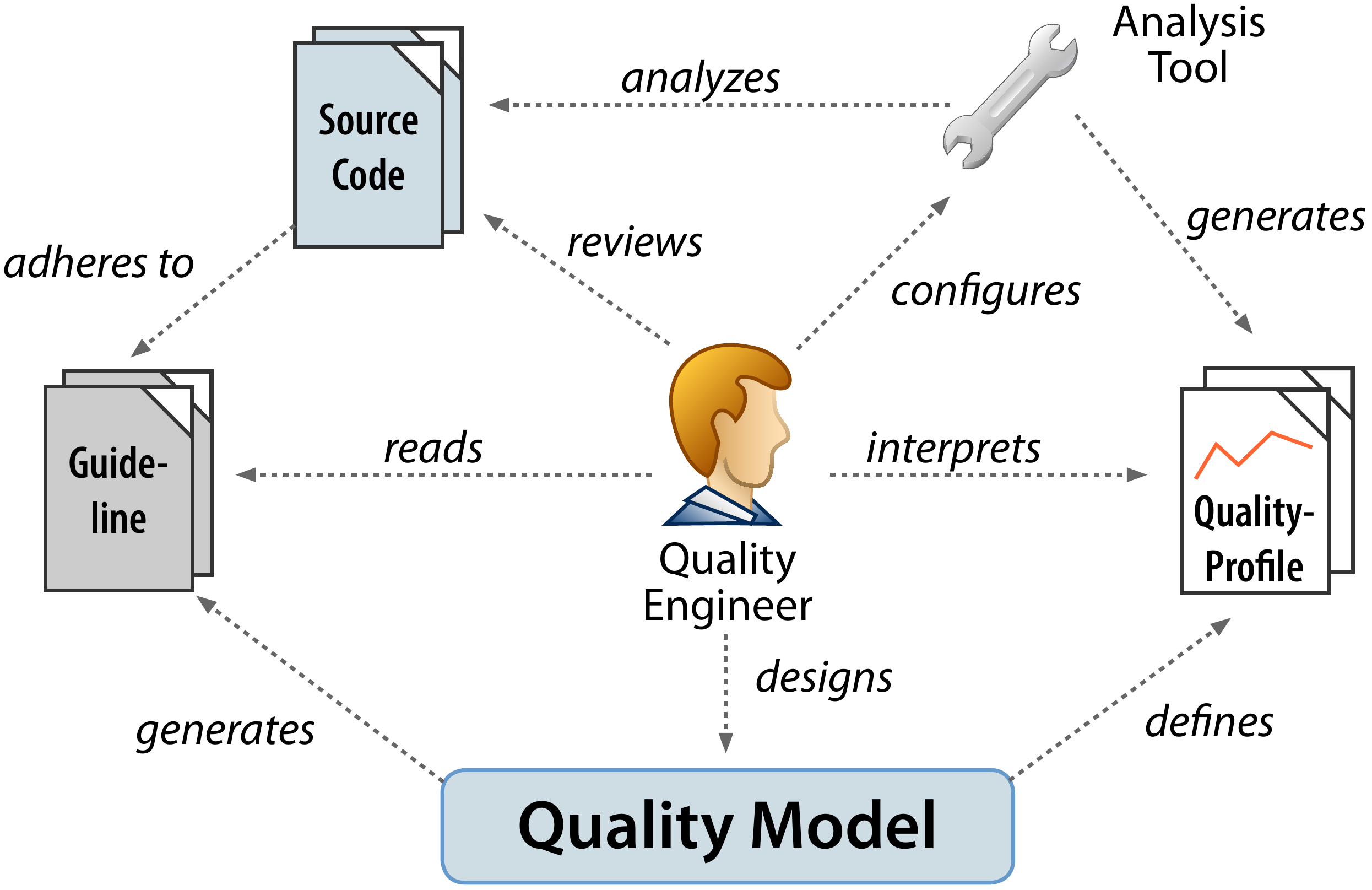}
\caption{Model-Based Quality Controlling\label{fig:controlling}}
\end{center}
\vspace{-1em}
\end{figure}

In contrast to other quality models that are expressed in terms of prose 
and graphics only, our maintainability model is truly integrated in the 
software development as  basis of all quality assurance activities. As 
Fig.~\ref{fig:controlling} shows, the model can be seen as project- or 
company-wide quality knowledge base that centrally stores the definition 
of quality in a given context. Of course, an experienced quality engineer 
is still needed for designing the quality models and enforcing them with 
manual review activities. However, he can rely on a single definition of 
quality and is supported by the automatic generation of guidelines. 
Moreover, quality assessment tools like static analyzers that automatically 
assess artifacts can be directly linked to the quality model and do not 
operate isolated from the centrally stored definition of quality. 
Consequently, the quality profiles generated by them are tailored to match 
the quality requirements defined in the model. We refer to this approach as 
\emph{model-based quality controlling}.

The following sections explain the basic concepts of our model and discuss 
the differences to classical hierarchical models.

\SubSection{Hierarchical Models}

\begin{figure}[h]
\begin{center}
\includegraphics{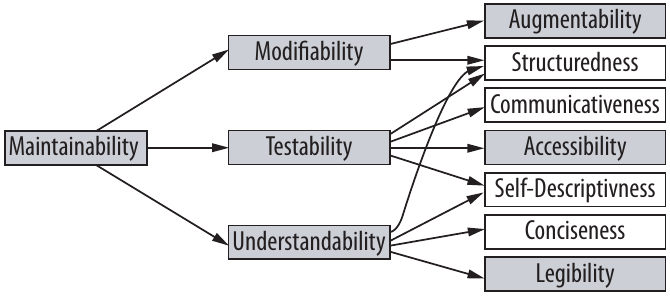}
\caption{Software Quality Tree\label{fig-boehm}}
\end{center}
\vspace{-1em}
\end{figure}

The idea of explicitly modeling maintenance activities was based 
on our experiences with building large hierarchical quality models. In turn 
of this process it became harder and harder to maintain a consistent model 
that adequately describes the interdependencies between the various quality 
criteria. A thorough analysis of this phenomenon revealed that our model 
and indeed most previous models mixed up nodes of two very different kinds: 
maintenance \emph{activities} and \emph{characteristics} of the system to 
maintain. An example for this problem is given in Fig.~\ref{fig-boehm} 
which shows the \emph{maintainability} branch of Boehm's \emph{Software 
Quality Characteristics Tree} \cite{1978_boehmb_software_quality}.
  
Though (substantivated) adjectives are used as descriptions, the nodes in 
the gray boxes refer to activities whereas the uncolored nodes describe 
system characteristics (albeit very general ones). So the model should 
rather read as: When we \emph{maintain} a system we need to \emph{modify} 
it and this activity of \emph{modification} is (in some way) influenced by 
the \emph{structuredness} of the system.  While this difference may not 
look important at first sight, we claim that this mixture of activities and 
characteristics is at the root of most problems encountered with previous 
models. The semantics of the edges of the tree is unclear or at least 
ambiguous because of this mixture. And since the edges do not have a clear 
meaning they neither indicate a sound explanation for the relation of two 
nodes nor can they be used to aggregate values. 

As the actual maintenance efforts strongly depend on both, the type of
system and the kind of maintenance activity, it should be obvious that the
need to distinguish between activities and characteristics becomes not only
clear but imperative. This can be illustrated by the example of two
development organizations where company $A$ is responsible for adding
functionality to a system while company $B$'s task is merely fixing bugs of
the same system just before its phase-out. One can imagine that the success
of company $A$ depends on different quality criteria (\eg architectural
characteristics) than company $B$'s (\eg a well-kept bug-tracking system).
While both organizations will pay attention to some common attributes such
as documentation, $A$ and $B$ would and should rate the maintainability of
$S$ in quite different ways because they are involved in fundamentally
different \emph{activities}.

Focusing on the individual factors that influence productivity within a 
certain context widens the scope of the relevant criteria. $A$ and $B$'s 
productivity is not only determined by the system itself but by a plethora 
of other factors which include the skills of the engineers, the presence of 
appropriate software processes and the availability of proper tools like 
debuggers. To clarify that our observations are not limited to the software 
system itself,  we refer to the \emph{situation} instead of the
\emph{software system} from now on. Similar to the software 
system that can be decomposed in \emph{components}, the situation can be 
decomposed in, what we call \emph{facts}.

\SubSection{An Activity-Based Model for Maintainability}

The consequent separation of activities and facts leads to a new 
2-dimensional quality model that regards \emph{activities} and \emph{facts} 
as first-class citizens for modeling maintainability.

\begin{figure}[h]
\begin{center}
\includegraphics{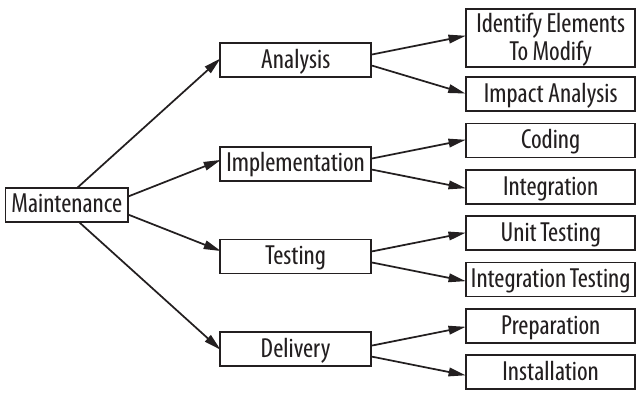}
\caption{Example Activities\label{fig-acts}}
\end{center}
\vspace{-1em}
\end{figure}

The set of relevant activities depends on the particular development and
maintenance process of the organization that uses the quality model. As an
example, we use the IEEE 1219 standard maintenance
process\,\cite{1998_ieee_standard_1219}.  Its activity breakdown
structure is depicted in Fig.~\ref{fig-acts}.  For the sake of brevity we
only show a subset of the activities. 

\begin{figure}[h]
\begin{center}
\includegraphics{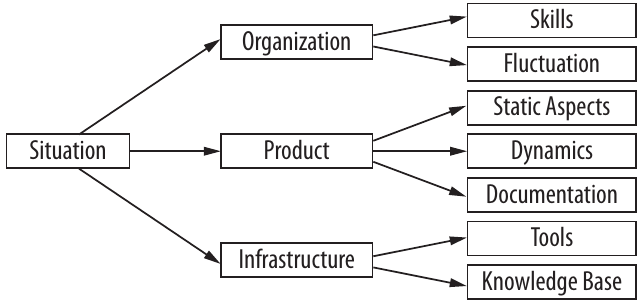}
\caption{Example Facts\label{fig-facts}}
\end{center}
\vspace{-1em}
\end{figure}

The 2\textsuperscript{nd} dimension of the model, the facts of the situation, are modeled
similar to an FCM model but without activity-based nodes like
\emph{augmentability}. It is important to understand, that we do not limit
this dimension to properties of the software system,
e.\,g.~\emph{structuredness}, but try to capture all factors that affect
one or more activities. An excerpt of a facts tree is shown in 
Fig.~\ref{fig-facts}. 

Obviously, the granularity of the facts shown in the diagrams are too 
coarse to be actually evaluated. We follow the FCM approach in the 
situation tree by breaking down high level facts into detailed, tangible 
ones which we call \emph{atomic} facts. An atomic fact is a fact that can 
or must be assessed without further decomposition either because its 
assessment is obvious or there is no known decomposition. 

\begin{figure}[h]
\begin{center}
\includegraphics{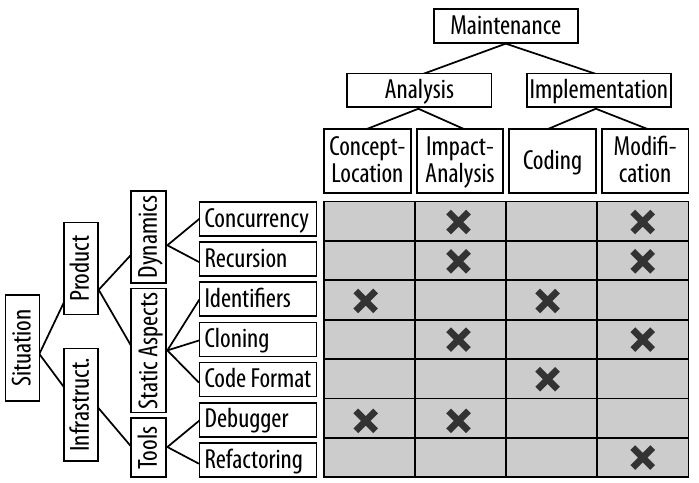}
\caption{Maintainability Matrix\label{fig-matrix}}
\end{center}
\vspace{-1em}
\end{figure}

To achieve or measure maintainability in a given project setting we now
need to establish the interrelation between facts and activities.  Because
of the tree-like structures of activities and facts it is sufficient to
link atomic facts with atomic activities. This relationship is best
expressed by a matrix as depicted in the simplified Fig.~\ref{fig-matrix}.

The matrix points out what activities are affected by which facts and 
allows to aggregate results from the atomic level onto higher levels in 
both trees because of the unambiguous semantics of the edges. So, one can 
determine that concept location is affected by the names of identifiers and 
the presence of a debugger. Vice versa, cloned code has an impact on two 
maintenance activities. The example depicted here uses a Boolean relation 
between facts and activities and therefore merely expresses the existence 
of a relation between a fact and an activity. To express different 
directions and strengths of the relations, more elaborate scales can be
used here. Below, we show the application of a three-valued scale that
proved to be sufficient for our current work.

The aggregation within the two trees provides a simple means to cross-check 
the integrity of the model.  For example, the sample model in 
Fig.~\ref{fig-matrix} states, that tools do not have an impact on coding, 
which is clearly nonsense. The problem lies in the incompleteness of the 
depicted model, that does not include tools like integrated development 
environments.

\SubSection{Attributes \& Impacts}

We found that a fine-granular decomposition of the situation (the facts 
tree) inevitably leads to a high number of repetitions as the same properties 
apply to different kind of artifacts. For example, \emph{consistency} is 
obviously required for identifier names as well as for the layout of the
documentation.

Therefore our model further decomposes facts into \emph{entities} and 
\emph{attributes} where entities ``are the objects we observe in the real 
world'' and attributes are ``the properties that an entity 
possesses''~\cite{1995_kitchemhamb_measurement}. Hence, entities describe a 
plain decomposition of the situation. Examples are \emph{documentation}, 
\emph{classes}, \emph{variables} or the available \emph{infrastructure}. 
Entities are associated with one or more attributes like 
\emph{consistency}, \emph{redundancy}, \emph{completeness} or 
\emph{superfluousness}.

So, the facts defined in the facts tree are actually  tuples of entities 
and attributes: \fact{Entity $\mathsf{e}$}{Attribute $\mathsf{a}$}. They 
describe properties of the situation that are desired or undesired in the 
context of maintainability.  Examples are \fact{Identifiers}{consistency}, 
\fact{Documentation}{completeness} or \fact{Debugger}{existence} that 
simply describes the presence or absence of a debugging tool.

Note that the separation of entities and attributes does not only reduce 
redundancy but allows for a clean decomposition of the situation. This can 
be illustrated by an example of the quality taxonomy defined 
in~\cite{1992_omanp_maintainability}: \emph{System Complexity}. As 
\emph{System Complexity} appears too coarse-grained to be assessed directly, 
is desirable to further decompose this element. However, the decomposition 
is difficult as the decomposition criterion is not clearly defined, \ie it 
is not clear what a subelement of \emph{System Complexity} is. A separation 
of the entity and the attribute as in \fact{System}{Complexity} allows for 
a cleaner decomposition as entities themselves are not valued and can
be broken up in a straightforward manner, \eg in
\fact{Subsystem}{Complexity} or \fact{Class}{Complexity}.

\paragraph{Impacts}

Using the notation introduced for facts we can elegantly express the impact 
a fact has on an activity with a three-valued scale where `$+$' expresses a 
positive and `$-$' a negative impact (the non-impact is usually not made 
explicit):

\begin{center}
\impact{Entity $\mathsf{e}$}{Attribute $\mathsf{a}$}
{Activity $\mathsf{a}$}{+/-}
\end{center}

Examples are \impactp{Debugger}{Existence}{Fault Diagnostics}, that 
describes that the existence of a debugger has a positive influence on the 
activity fault diagnostics. \impactp{Identifiers}{Consistency}{Concept 
Location} describes that consistently used identifier names have a positive 
impact on the concept location activity. 
\impactn{Variable}{superfluousness}{Code Reading} describes that unused 
variables hamper the reading of the code. To overcome the problem of
unjustified quality guidelines each impact is additionally equipped with
a detailed description. 

\paragraph{Assessment}

Obviously, the facts are the elements of the model that need to be assessed 
in order to determine the maintainability (or maintenance effort) of a 
situation. Since many important facts are semantic in nature and inherently 
not assessable in an automatic manner, we carefully distinguish three fact 
categories:

\begin{enumerate}

\item Facts that can be assessed or measured with a tool. An example is an 
automated check for \texttt{switch}-statements without a 
\texttt{default}-case (\fact{Switch Statement}{completeness}).

\item Facts that require manual activities; e.\,g.~reviews. An example is a 
review activity that identifies the improper use of data structures 
(\fact{Data Structures}{appropriateness}).

\item Facts that can be automatically assessed to a limited extent 
requiring additional manual inspection. An example is redundancy analysis 
where cloned source code can be found with a tool but other kinds of 
redundancy must be left to manual inspection (\fact{Source 
Code}{redundancy}).
  
\end{enumerate}

\SubSection{The Quality Metamodel}

\begin{figure}[h]
\begin{center}
\includegraphics[width=0.99\linewidth]{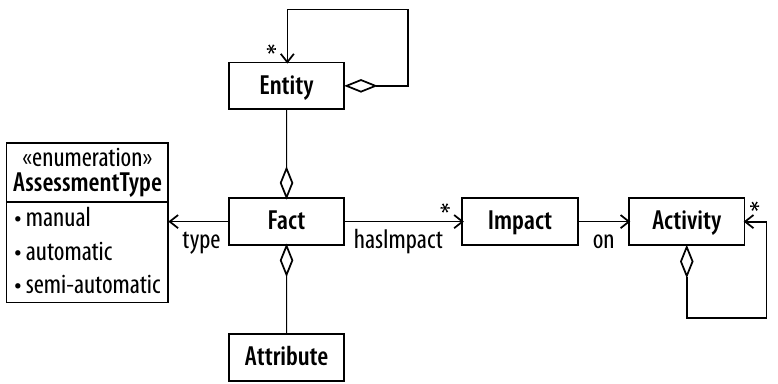}
\caption{The Quality Metamodel QMM\label{fig-qmm}}
\end{center}
\vspace{-1em}
\end{figure}

Although most quality models conform to an implicitly defined metamodel 
they usually lack an explicitly specified metamodel that precisely defines 
the set of legal model instances. In contrast to this, our model is based 
on the explicit quality metamodel \qmm. This metamodel consists of the 
elements discussed above: entities, attributes, facts, activities and 
impacts. For space reasons we do not explain this metamodel in detail but 
present a UML class diagram that illustrates the different model elements 
and their interplay (Fig.~\ref{fig-qmm}). Please note, that the figure 
shows only the core elements and omits details like the explanation texts 
that are associated with each element. Moreover, it does not show that the 
model features a generalization mechanism that allows attribute 
inheritance. It is, for example, possible to specify an attribute 
\attribute{superfluousness} for the entity \entity{Component} and inherit it
to the entity \entity{Class}. 

The benefit of an explicit metamodel is twofold: First, it ensures a
consistent structure of quality models. Second, it is a necessary basis
for modeling tool support as described in the next section.

\SubSection{Tool Support\label{sec:tools}}

Comprehensive maintainability models typically contain several hundred 
model elements. For example, the model that was developed for a commercial 
project in the field of telecommunication~\cite{2006_broym_maintainability} has 
a total of 413 model elements consisting of 160 facts (142 entities and 16 
attributes), 27 activities and 226 impacts. Hence, quality models demand  a 
rich tool set for their efficient creation, management and application just 
like other large models, \eg UML class diagrams.

Due to the fact that our quality models are based on an explicit metamodel 
we are able to provide a model editor that does not only allow the initial 
development of quality models but also supports other common tasks like 
browsing, persistence, versioning and refactoring\footnote{A beta version 
of the editor can be downloaded from 
\\\url{http://www4.cs.tum.edu/~ccsm/qmm/}}.

One of the most powerful features of the model editor is the automatic 
generation of guideline documents from the quality model. This enables us 
to transfer the abstract definition of quality stored in the model to a 
format developers are familiar with. However, unlike classic, hand-written 
guidelines the automatically generated ones are guaranteed to be 
synchronized with the quality model that explicitly captures the 
understanding of quality within a project or a company. Guideline documents 
can be tailored to specific needs by defining selected views on the model. 
For example, a guideline document could be specifically generated to be 
used during documentation review sessions.

\SubSection{Summary}

Our approach to modeling maintainability is based on the quality metamodel 
\qmm. It advances on previous approaches to model maintainability  and 
quality with respect to the following issues:

\paragraph{Focus on Activities} Our model enforces a clear separation 
of system characteristics and maintenance activities. This separation makes 
activities, which constitute the main cost factor in software maintenance, 
first-class citizens in quality modeling and thereby contributes to a 
discussion of quality economics~\cite{2007_wagerns_quality_economics}.

\paragraph{Unambiguous Decomposition Criteria} Previous hierarchical models 
often exhibit a ``somewhat arbitrary selection of characteristics and 
sub-characteristics''~\cite{1997_kitchenhamb_squid_quality_model, 
1996_kitchenhamb_quality_elusive_target}. We claim this is due to the fact 
that previous models typically lack a clearly defined decomposition 
criterion. For example, it is not entirely clear how 
\emph{self-descriptiveness} relates to \emph{testability} in Boehm's 
quality characteristics tree if one is not satisfied with a trivial ``has 
something to do with''. 
Our approach overcomes this shortcoming by rigorously separating aspects 
that are typically intermingled: activities, entities and attributes. This
separation creates separate hierarchies with clearly defined decomposition
criteria.

\paragraph{Scope} Our approach is not limited to modeling quality 
characteristics of a system itself. It includes external but equally 
important organizational issues like the existence of a configuration 
management process or the available tool infrastructure.

\paragraph{Explicit Metamodel} Unlike other approaches known to us, our 
approach is based on an explicitly defined metamodel. This enables us to 
provide a rich set of tools for editing and maintaining quality models.
Most importantly, the metamodel is a key required for the model-based
quality controlling approach outlined before.
Additionally, the metamodel fosters the conciseness, consistency and
completeness of quality models as it forces the model designer to stick
to an established framework and supports him in finding omissions. Examples
are given in the next section.

\Section{Case Study}
\label{sec:case}

The applicability and usefulness of the approach described above was 
evaluated in a case study with a German truck and bus manufacturer. 
Specifically, we built a model for the maintainability of the Matlab 
Simulink and Stateflow models used for code generation. For this, 
the quality model we had developed in a project in the field of 
telecommunication was modified and extended with Simulink/Stateflow-specific
elements. Guideline documents were generated from the model and 
automatic analyses were derived.

\SubSection{Environment}

\paragraph{MAN Nutzfahrzeuge Group}
The MAN Nutzfahrzeuge Group is a German-based international supplier of
commercial vehicles and transport systems, mainly trucks and busses. It has
over 34,000 employees world-wide of which 150 work on electronics and
software development. Hence, the focus is on embedded systems in the
automotive domain. 

The organization brought its development process to a
high level of maturity by investing enough effort to redesign it according
to best practices and safety-critical system standards. The driving force
behind this redesign was constantly focusing on how each activity
contributes to global reliability and effectiveness. Most parts of the
process are supported by an integrated data backbone developed on the eASEE
framework from Vector Consulting GmbH.  On top of this backbone, a complete
model-based development approach has been established using the tool chain
of Matlab/Simulink and Stateflow as modeling and simulation environment and
TargetLink of dSpace as C-code generator. We describe the application and
adoption of our model to this concrete situation and the generated
benefits. The study lead to the adoption of the model into the MAN standard
development process.

\paragraph{Embedded Systems and Matlab/Simulink}

Matlab/Simulink is a model-based development suite aiming at the
embedded systems domain. It is commonly used in the automotive
area. The original \emph{Simulink} has its focus on continuous control
engineering. Its counterpart \emph{Stateflow} is a dialect of statecharts
that is used to model the event-driven parts of a system. The Simulink
environment already allows to simulate the model in order to validate it.

In conjunction with code generators such as Embedded Coder from
MathWorks or TargetLink by dSpace it enables the complete and
automatic transformation
of models to executable code. 
This is a slightly different flavor of model-based development
than the MDA approach
proposed by the OMG\footnote{\url{http://www.omg.org/mda/}}. 
There is no explicit need to have different
types of models on different levels and the modeling language is not
UML. Nevertheless, many characteristics are similar and quality-related
results could easily be transferred to an MDA setting.

\SubSection{The Maintainability Model}

The initial maintainability model that was developed in the field of 
telecommunication (Sec.~\ref{sec:qmm}) already covered various areas that we 
consider important for MAN, too. Examples are the parts of the 
model dedicated to architectural aspects or to the development process. 

In the case study, we augmented the existing maintainability model with 
model elements that address Simulink/Stateflow-models which are used as 
basis for code generation. Although such models are seemingly different 
from traditional source code, we found that a great number of 
source-code-related facts could be reused for them as they fundamentally serve the same 
aim: specify executable production-code. 

Specifically, we extended the facts tree of the maintainability model with 
87 facts (64 new entities and 3 new attributes) that describe properties of 
entities not found in classical code-based developed. Examples are states, 
signals, ports and entities that describe the graphical representation of 
models, \eg colors. Furthermore, we modified the activities tree to match 
the MAN development process and added two activities (\emph{Model Reading} 
and \emph{Code Generation}) that are specific for the model-based 
development approach. 84 impacts describe the relation between facts and 
activities.

The newly developed parts of the maintainability model are based on three 
types of sources:  (1) existing guidelines for Simulink/Stateflow, (2) 
scientific studies about model-based development and (3) expert know-how of 
MAN's engineers.

Specifically, our focus lies on the consolidation of four guidelines 
available for using Simulink and Stateflow in the development of embedded 
systems: the MathWorks documentation \cite{mathworks06}, the MAN-internal 
guideline, the guideline provided by dSpace \cite{dspace06}, the developers 
of the TargetLink code-generator, and the guidelines published by the 
MathWorks Automotive Advisory Board (MAAB) \cite{2001_maab_guideline}.

Because of space and confidentiality reasons, we are not able to fully 
describe the MAN-specific model here. However, we present a number of 
convincing examples that demonstrate how our approach helps to overcome 
different kinds of shortcomings.

We start with a simple translation of the existing MAN guidelines for 
Stateflow models into the maintainability model. For example, the MAN 
guideline requires the current state of a Stateflow chart to be available 
as an measurable output. This simplifies testing of the model and improves 
the debugging process. In terms of the model this is expressed as 
\impactp{Stateflow Chart}{Accessibility}{Debugging} and \impactp{Stateflow 
Chart}{Accessibility}{Test}.

We describe the ability to determine the current state with the attribute 
\attribute{accessibility} of the entity \entity{Stateflow Chart}.  The 
Stateflow chart contains all information about the actual statechart model. 
Note that we carefully distinguish between the \emph{chart} and the 
\emph{diagram} that describes the graphical representation. In the model 
the facts and impacts have additional fields that describe the relationship 
in more detail. This descriptions are included in generated guideline 
documents.

\paragraph{Consolidation of the Terminology}

In the case study we found that building a comprehensive quality model has 
the beneficial \emph{side-effect} of creating a consistent terminology. By 
consolidating the various sources of guidelines, we discovered a very 
inconsistent terminology that hampers a quick understanding of the 
guidelines. Moreover, we found that even at MAN the terminology has not 
been completely fixed. Fortunately, building a quality model automatically 
forces the modeler to give all entities explicit and consistent names. The 
entities of the facts tree of our maintainability model automatically 
define a consistent terminology and thereby provide a glossary.

One of many examples is the term \emph{subsystem} that is used in the 
Simulink documentation to describe Simulink's central means of 
decomposition. The dSpace guideline, however, uses the same term to refer 
to a \emph{TargetLink subsystem} that is similar to a Simulink subsystem 
but has a number of additional constraints and properties defined by the 
C-code generator. MAN engineers on the other hand, usually refer to a 
\emph{TargetLink subsystem} as \emph{TargetLink function} or simply 
\emph{function}. While building the maintainability model, this discrepancy
was made explicit and could be resolved. 

\paragraph{Resolution of Inconsistencies}
Furthermore, we are not only able to identify inconsistencies in the terminology
but also in contents. For the entity \emph{Implicit Event} we found completely
contradictory statements in the MathWorks documentation and the dSpace
guidelines.
\begin{itemize}
\item \emph{MathWorks \cite{mathworks06}} ``Implicit event broadcasts [\ldots] and implicit
conditions [\ldots] make the diagram easy to read and the generated code more
 efficient.''
\item \emph{dSpace \cite{dspace06}} ``The usage of implicit events is therefore intransparent
concerning potential side effects of variable assignments or the entering/exiting 
of states.''
\end{itemize}
Hence, MathWorks sees implicit events as improving the readability while
dSpace calls them intransparent. This is a clear inconsistency. After discussing
with the MAN engineers, we adopted the dSpace view.

\paragraph{Revelation of Omissions}

An important feature of the quality metamodel is that it supports 
inheritance. This became obvious in the case study after modeling the MAN 
guidelines for Simulink variables and Stateflow variables. We model them 
with the common parent entity \entity{Variable} that has the attribute 
\attribute{Locality} that expresses that variables must have the smallest 
possible scope. As this attribute is inherited by both types of variables, 
we found that this important property is not expressed in the original 
guideline. Moreover, we see by modeling that there was an imbalance between 
the Simulink and Stateflow variables. Most of the guidelines related only 
to Simulink variables. Hence, we transferred them to Stateflow as well.

\paragraph{Integration of Recent Research Results}
Finally, we give an example of how a scientific result can be incorporated
into the model to make use of new empirical research. The use of Simulink
and Stateflow has not been intensively investigated in terms of maintainability.
However, especially the close relationship between Stateflow and the UML
statecharts allows to reuse results. A study on hierarchical states in
UML statecharts \cite{cruz05} showed that the use of hierarchies improves the
efficiency of understanding the model in case the reader has a certain amount
of experience. This is expressed in the model as follows:
\impactp{Stateflow Diagram}{Structuredness}{Model reading}.

\SubSection{Usage of the Model}
\vspace{-1em}
In the case study, we concentrated on checklist generation and some 
preliminary automatic analyses. Those were chosen because they promised the 
highest immediate pay-off.
\vspace{-1em}

\paragraph{Checklist Generation} We see quality models as central knowledge 
bases w.r.t.~quality issues in a project, company, or domain. This 
knowledge can and must be used to guide development activities as well as 
reviews. However, the model in its totality is too complex to be 
comprehended entirely. Hence, it cannot be used as a quick reference. 
Therefore, we exploit the tool support for the quality model to select subsets 
of the model and generate concise guidelines and checklists for specific 
purposes.

Automatic generation of guideline documents was perceived to be highly 
valuable as the documents could be structured to be read conveniently by 
novices as well as experts. Therefore the documents feature a very compact 
checklist-style section with essential information only. This 
representation is favored by experts who want to ensure that they comply 
to the guideline but do not need any further explanation. For novices the 
remainder of the document contains a hyperlinked section providing 
additional detail. Automatic generation enables us to conveniently change the 
structure of all generated documents. More importantly, it ensures consistency 
within the document which would be error-prone in hand-written documents.

\paragraph{Preliminary Automatic Analyses.} As the model is aimed at 
breaking down facts to a level where they can be assessed and they are 
annotated with the degree of possible automation, it is straightforward to 
implement automatic analyses. So far, we have not fully exploited the 
possibilities but we are able to show that simple facts can be checked in 
Simulink and Stateflow models. For this, we wrote a parser for the 
proprietary text format used by Matlab to store the models. Using this 
parser we are able to determine basic size and complexity metrics of model 
elements like states, blocks, etc. Moreover, we can use the parser to 
automatically identify model elements that are not satisfactorily supported 
by the C-code generator. By integrating these analyses in our quality 
controlling toolkit \textsc{ConQAT}~~\cite{2005_deissenboeckf_conqat} we
are able to create aggregated quality profiles and powerful visualizations
of quality data.

\SubSection{Discussion}
\vspace{-1em}
The metamodel and the corresponding method for modeling maintainability 
proposed in Sec.~\ref{sec:qmm} proved to be applicable to industrial development 
environments in the case study. After a short time, the 2-dimensional 
structure was accepted by the MAN engineers. Especially the model's explicit 
illustration of impacts on activities was seen as beneficial as it provides 
a sound justification for the quality rules expressed by the model. Moreover, 
the general method of modeling -- that inherently includes structuring --
improved the guidelines: although the initial MAN guideline included many 
important aspects, we still were able to reveal several omissions and 
inconsistencies. Building the model, similar to other model building 
activities in software engineering \cite{pretschner05}, revealed these 
problems and allowed to solve them.

Another important result is that the maintainability model contains a 
consolidated terminology. By combining several available guidelines, we 
could incorporate the quality knowledge contained in them and form
a single terminology. We found terms used consistently as well as 
inconsistent terminology. This terminology and combined knowledge base was 
conceived useful by the MAN engineers.

Although the theoretical idea of using an explicit quality metamodel for 
centrally defining quality requirements is interesting for MAN, the main 
interest is in the practical use of the model. For this, the generation of 
purpose-specific guidelines was convincing. We not only build a model to 
structure the quality knowledge but we are able to communicate that 
knowledge in a concise way to developers, reviewers and testers. Finally, the 
improved efficiency gained by automating specific assessments was seen as 
important. The basis and justification for these checks is given by the 
model.

\Section{Conclusion}
\vspace{-1em}
Although maintainability is undisputedly considered one of the fundamental 
quality attributes of software systems, the research community has not yet 
produced a sound and accepted definition or even a common understanding 
what maintainability actually is. Substantiated by various examples we 
showed that this shortcoming is due to intrinsic flaws of current 
approaches to define, assess and improve maintainability. We showed that 
there is a need to make maintenance activities  first-class citizens in 
modeling maintainability due to their economical importance. This notion is 
captured by our 2-dimensional quality metamodel which maps facts about a 
development situation to maintenance activities and thereby highlights 
their impact on the maintenance effort. 

In a case study in the automotive domain we showed that our metamodel and 
the accompanying tools could be successfully used to build a comprehensive 
maintainability model for the development of embedded systems with 
Simulink/Stateflow. The construction of the model helped to define a 
consistent terminology and to reveal omissions as well as contradictions in 
existing quality guidelines. Long-term benefits are gained by the automatic 
generation of specifically-tailored guideline documents and the usage of 
automatic quality assessments. The study lead to the inclusion 
of the model into the MAN standard development process.

Our future work with MAN focuses on widening the scope of the automated quality 
assessments. 
After first encouraging results with modeling 
\emph{usability}~\cite{2007_winters_usability}, we currently use the quality 
metamodel to model other quality attributes like reliability and performance. 
Furthermore, we plan to use an integrated quality model for all relevant quality 
attributes. Our aim is to unify the currently used isolated approaches to 
quality to enable a holistic but systematic discussion of quality. We are 
convinced that this is an important step towards our final goal of a truly 
economically justified practice of quality engineering.

\vspace{-1em}
\bibliographystyle{latex8}
\bibliography{icsm07}

\copyright 2007 IEEE. Personal use of this material is permitted. Permission from IEEE must be obtained for all other users,
including reprinting/republishing this material for advertising or promotional purposes, creating new collective works for resale or
redistribution to servers or lists, or reuse of any copyrighted components of this work in other works.

\end{document}